# Direct Experimental Evidence for Substrate Adatom Incorporation into a Molecular Overlayer

Philip J. Mousley[1], Luke A. Rochford[2], Paul T.P. Ryan[1,3], Philip Blowey[1,4], James Lawrence[5], David A. Duncan[1], Hadeel Hussain[1], Billal Sohail[5], Tien-Lin Lee[1], Gavin R. Bell[4], Giovanni Costantini[5], Reinhard J. Maurer[5], Christopher Nicklin[1], D. Phil Woodruff[4*]

[1]*Diamond Light Source, Harwell Science and Innovation Campus, Didcot, OX11 0DE,*
[2]*Chemistry Department, University of Birmingham, University Road, Birmingham B15 2TT, UK*
[3]*Department of Materials, Imperial College, London SW7 2AZ, UK*
[4]*Department of Physics, University of Warwick, Coventry CV4 7AL, UK*
[5]*Department of Chemistry, University of Warwick, Coventry CV4 7AL, UK*

## ABSTRACT

While the phenomenon of metal substrate adatom incorporation into molecular overlayers is generally believed to occur in several systems, the experimental evidence for this relies on the interpretation of scanning tunnelling microscopy (STM) images, which can be ambiguous and provides no quantitative structural information. We show that surface X-ray diffraction (SXRD) uniquely provides unambiguous identification of these metal adatoms. We present the results of a detailed structural study of the Au(111)-$F_4$TCNQ system, combining surface characterisation by STM, low energy electron diffraction and soft X-ray photoelectron spectroscopy with quantitative experimental structural information from normal incidence X-ray standing waves (NIXSW) and SXRD, together with dispersion corrected density functional theory (DFT) calculations. Excellent agreement is found between the NIXSW data and the DFT calculations regarding the height and conformation of the adsorbed molecule, which has a twisted geometry rather than the previously supposed inverted bowl shape. SXRD measurements provide unequivocal evidence for the presence and location of Au adatoms, while the DFT calculations show this reconstruction to be strongly energetically favoured.

---

[*] Email: <u>d.p.woodruff@warwick.ac.uk</u>

**INTRODUCTION**

It is now recognised that molecular adsorption on metal surfaces often leads to significant modification of the structure of both the adsorbed molecule and the metal surface. One example of this metal surface modification is the reported adsorption-induced incorporation of metal adatoms from the bulk into the molecular overlayer, often forming two-dimensional metal organic frameworks (*e.g.* [1]). This effect is also believed to play a key role in surface-assisted Ullmann coupling reactions (*e.g.* [2]). Despite several reports of this phenomenon, there are no quantitative experimental determinations of these surface structures. The present evidence for the phenomenon comes mostly in the form of atomic-scale protrusions seen in constant tunnelling current scanning tunnelling microscopy (STM) images, interpreted as being due to these metal adatoms. However, STM provides no identification of the atomic species leading to such features, and indeed there is even no totally reliable correlation between atomic scale protrusions seen in STM images and the positions of surface atoms.[3, 4] In a few studies, the assignment of these features to metal adatoms is supported by density functional theory (DFT) simulations of the STM images, based on the simple Tersoff-Hamman approach,[5] which takes no account of the role of the tunnelling tip in the imaging process.

In recent years, the experimental technique that has provided most of the quantitative information on the structure of metal-organic interfaces, and particularly on the height of the adsorbed molecule above the surface, is normal incidence X-ray standing waves (NIXSW).[6] This technique generates quantitative information on the location of the different constituent atoms of an adsorbed molecule relative to the underlying substrate, by monitoring their core level photoemission as the X-ray standing wave, established at a Bragg reflection, is swept through the crystal. This information is element specific due to the characteristic photoelectron binding energies of these core levels, while chemical shifts in these energies makes it also possible to determine the distinct local sites of atoms of the same element in different chemical bonding states within the molecule. However, NIXSW is not able to distinguish between emission from metal adatoms and from the vastly larger number of atoms of the same metal in the underlying substrate; any associated chemical shifts are too small to be exploited, even using detection at grazing emission angles. The technique is thus 'blind' to metal adatoms, although their presence may be inferred from their impact on the resulting molecular conformation, as in the case of 7,7,8,8-tetracyanoquinodimethane (TCNQ) adsorbed on Ag(111).[7]

One technique that can be expected to provide direct evidence of the presence and location of metal adatoms is surface X-ray diffraction (SXRD). Although the general technique of XRD is not explicitly element-specific, the scattering cross-sections scale as the square of the atomic number, $Z$. One consequence of this is that SXRD is regarded as unsuitable to determine the structure of adsorbed molecules in which all the constituent atoms have low $Z$ (C, N, O, H). However, if the underlying metal has a much higher $Z$ (*e.g.* Cu, Ag, Au), metal adatom incorporation will lead to the intensities of diffracted beams arising from the overlayer periodicity being dominated by scattering from the adatoms in this layer. This sensitivity to the location of high-$Z$ elements is the basis of the 'heavy atom' method of solving complex macromolecular crystal structures with XRD and is also relevant to the related techniques of isomorphous replacement and multiwavelength anomalous dispersion XRD (MAD) (*e.g.* [8]).

Here, we present the results of an investigation to demonstrate and explore the application of SXRD to a 2D metal-organic overlayer believed to be created by metal adatoms. The model system we have chosen to investigate is the fully fluorinated version of TCNQ, $F_4$-TCNQ, adsorbed on Au(111),[9] for which published STM results, supported by DFT image simulations, have been interpreted as evidence for Au adatom incorporation into the molecular overlayer. There have also been both earlier[10] and later[11] reports of DFT calculations of $F_4$TCNQ adsorption on Au(111), but these take no account of any possible surface reconstruction. The underlying motivation for investigating this, and closely related adsorption structures, is the need to understand the critical role that metal-organic interfaces play in determining the electronic properties of organic devices. TCNQ and $F_4$TCNQ have attracted considerable interest as additives in organic electronics, due to their strong electron acceptor properties, resulting in several model studies of these species adsorbed on coinage metal surfaces (*e.g.*[7,9--17]), while $F_4$TCNQ adsorption and incorporation into Ag surfaces has been shown to produce high work function electrodes for organometallic growth.[18]

Our comprehensive experimental and theoretical investigation of the ordered Au(111)-$F_4$TCNQ adsorption phase is based on initial characterisation with STM, low energy electron diffraction (LEED), and X-ray photoelectron spectroscopy (XPS) using incident synchrotron radiation at a photon energy of 2.2 keV. Complementary quantitative structural information has then been obtained from NIXSW and SXRD. We have also undertaken a new investigation of the energetics and structure through dispersion-corrected DFT calculations. Although the earlier investigation of this system based on STM[9] included some DFT calculations, these did not contain any corrections for van der Waals interactions that are known to strongly influence the adsorption height of these molecules on metal surfaces. Moreover, this earlier publication reported no quantitative structural results. Our results provide a completely consistent experimental and theoretical picture of the structure of $F_4$TCNQ adsorption on Au(111), the SXRD measurements providing unequivocal experimental evidence for the presence and location of Au adatoms in the overlayer.

**METHODS**

**Experimental Details**

Initial characterisation of the conditions for the preparation of the ordered adsorption phase of $F_4$TCNQ on Au(111) was performed in a combined LEED/STM ultra-high vacuum (UHV) surface science chamber at the University of Warwick. The Au(111) sample was subjected to *in situ* cleaning by cycles of argon ion bombardment and annealing to achieve constant current STM images and LEED patterns showing the characteristic 'herring-bone' reconstruction of the clean surface, which is lifted by adsorption of $F_4$TCNQ. All imaging and LEED patterns were obtained at room temperature, while the LEED patterns were recorded using a low incident beam current microchannel-plate amplified MCP-LEED optics. All STM images were plane corrected and flattened using the open source image-processing software Gwyddion.[19]

Further characterisation of the surface by SXPS, together with quantitative structural information from NIXSW (also recorded at room temperature), was obtained using the UHV surface science endstation at beamline I09 of the Diamond Light Source.[20] Sample cleaning and F$_4$TCNQ deposition using an organic molecular beam epitaxy source was performed using the same methods of the initial characterisation as at the University of Warwick, a similar MCP-LEED optics providing a direct cross-reference of the successful formation of the $\begin{pmatrix} 5 & 2 \\ 1 & 3 \end{pmatrix}$ ordered F$_4$TCNQ adsorption phase. All measurements were made using the crystal monochromator branch of this beamline that delivers 'hard' X-rays with energies greater than ~ 2keV. The endstation chamber is equipped with a VG Scienta EW4000 concentric hemispherical electron energy analyser with an extra-wide (±30°) angle acceptance mounted at 90° to the incident beam, which is used to collect core level photoemission spectra for both of these techniques.

NIXSW measurements were taken by stepping the photon energy of the X-ray beam at normal incidence to the surface through the (111) Bragg reflection of the Au substrate at a nominal energy of 2636 eV, recording the photoemission spectra around the C 1s, N 1s and F 1s emission at each step. These spectra were fitted by the chemically-shifted components and the variation of intensity of each components as a function of photon energy was then fitted by the standard NIXSW formulae (taking account of the influence of backward-forward asymmetry in the photoemission angular dependence)[6] to yield optimum values for the two key structural parameters, the coherent fraction, $f$, and the coherent position, $p$. Non-dipolar effects in the angular dependence of the high-energy photoemission were corrected as previously described[6] using values of the backward-forward asymmetry parameter, $Q$, derived from the published theoretical calculations[21]. It was assumed that these non-dipolar effects in the measurements using a wide angular range of emission detection could be modelled by a mean value of the polar emission angle, $\theta$, defined as the angle between the photon polarisation and the photoelectron detection direction. Due to the strongly attenuated signal coming from near 90° grazing emission angles, a $\theta$ of 18° was used.

SXRD measurements were made using the UHV surface science endstation of beamline I07 of the Diamond Light Source.[22] Sample preparation followed the same methods used at the University of Warwick and at beamline I09 and the formation of the required $F_4TCNQ$ ordered overlayer phase was checked with a standard LEED optics. An incident photon energy of 11.4 keV was used, chosen to be below the L-edges of Au, thereby avoiding a significant fluorescence background from the substrate, with a grazing incidence angle of 0.3°.

**DFT Calculations**

DFT calculations were performed with the FHI-aims package[23] and a GGA-PBE functional[24] was used to evaluate exchange-correlation. Dispersion interactions were modelled using the Tkatchenko-Scheffler $vdW^{surf}$ method (PBE+$vdW^{surf}$).[25] The adsorption structure was modelled as a periodically repeated cell comprising a single $\begin{pmatrix} 5 & 2 \\ 1 & 3 \end{pmatrix}$ unit mesh on Au(111) containing a single $F_4TCNQ$ molecule and either one Au adatom or no adatoms. The Au(111) surface was modelled as a slab consisting of four atomic layers and separated from its periodic image by a vacuum gap exceeding 60 Å. The coordinates of the atoms in the bottom two layers of the Au slab were constrained to the bulk truncated structure of Au and the positions of all other atoms in the simulation cell were relaxed. During optimisation we neglected long-range dispersion interactions between Au atoms and use the default "tight" basis set definition within FHI-aims. The Brillouin zone was sampled with an 8×8×1 Monkhorst-Pack[26] k-grid and the geometries were optimised to below a force threshold of 0.025 eV Å$^{-1}$. A dipole correction was employed in all cases. All DFT calculation inputs and outputs are freely available and can be found as a dataset in the NOMAD repository via https://dx.doi.org/10.17172/NOMAD/2022.01.31-1

**RESULTS AND DISCUSSION**

**Experimental Surface Characterisation**

Deposition of F$_4$TCNQ onto the Au(111) surface held at room temperature, from an organic molecular beam deposition source, led to the formation of the $\begin{pmatrix} 5 & 2 \\ 1 & 3 \end{pmatrix}$ ordered phase previously identified by STM[1] and LEED[27]. Figure 1(a) shows the resulting LEED pattern, obtained using a low incident current multichannel-plate amplified optics (MCP-LEED). Due to the low symmetry of the unit mesh of this phase, the pattern is a sum of the patterns from multiple domains related by the rotational and mirror symmetry elements of the substrate. Figure 1(b) shows a simulation, using the LEEDpat program[28] of the LEED pattern to be expected for the $\begin{pmatrix} 5 & 2 \\ 1 & 3 \end{pmatrix}$ mesh, with diffracted beams from different domains shown in different colours. The agreement with the experimentally recorded pattern is excellent confirming the validity of the unit cell assignment. Figure 1(c) shows a typical constant tunnelling current STM image of this surface. The spatial resolution of this STM image, recorded at room temperature, is undoubtedly inferior to the exceptional resolution of one of the images presented in the earlier paper by Faraggi *et al.*[9] that was recorded at 5 K, but the same periodicity is clear. In a very few images, such as Figure 1(d), there is possible evidence of the protrusions (one is circled) attributed in this earlier study to the presence of Au adatoms.

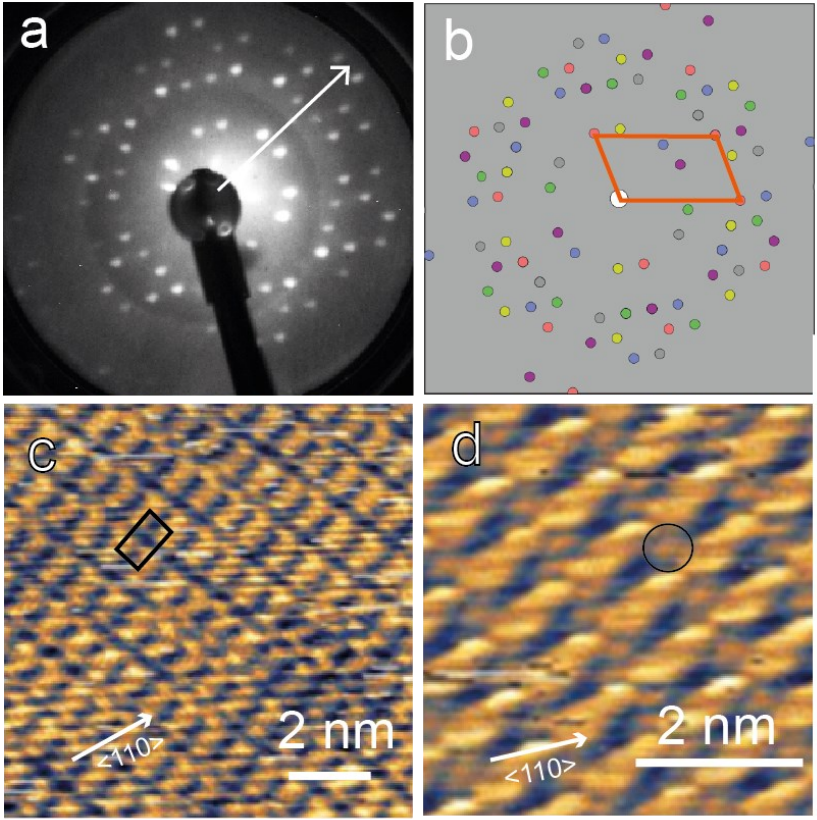

Figure 1. (a) LEED pattern recorded from the Au(111)-F$_4$TCNQ surface at an electron energy of 24.5 eV. (b) Simulation of the LEED pattern based on the $\begin{pmatrix} 5 & 2 \\ 1 & 3 \end{pmatrix}$ matrix, with diffracted beams from all rotational and mirror reflection domains shown in different colours. The reciprocal unit mesh of one of these (orange) is superimposed. (c) 10 nm × 10 nm constant current STM image of this surface with a unit mesh superimposed. (d) 5 nm × 5nm STM image of a different area showing some evidence (circled) of features previously attributed to Au adatoms. The arrows correspond to a <110> direction in the surface. STM imaging conditions, (sample bias and current): (c) +1.25 V, 250 pA; (d) -1.00 V, 75 pA).

Figure 2 shows XP spectra in the energy ranges of the C 1s, N 1s and F 1s emissions; the C 1s spectrum shows the chemically shifted components associated with CF, CC, and CN bonding, although the CN component is not clearly resolved in the raw spectrum. This spectrum is similar to one reported by Hählen *et al.* from a nominal monolayer of F$_4$TCNQ on Au(111),[29] and quite different from the spectrum resulting from multilayer deposition presented by these authors. The N 1s and F 1s spectra are fitted by single symmetric peaks, although the F 1s peak is significantly broader, possibly consistent with the presence of more than one unresolved component. The poorer statistics of the N 1s spectrum are due to its low photoionisation cross-section and the high background of the inelastic 'tail' of the intense Au 4d emission; longer data collection times were avoided to minimise radiation damage.

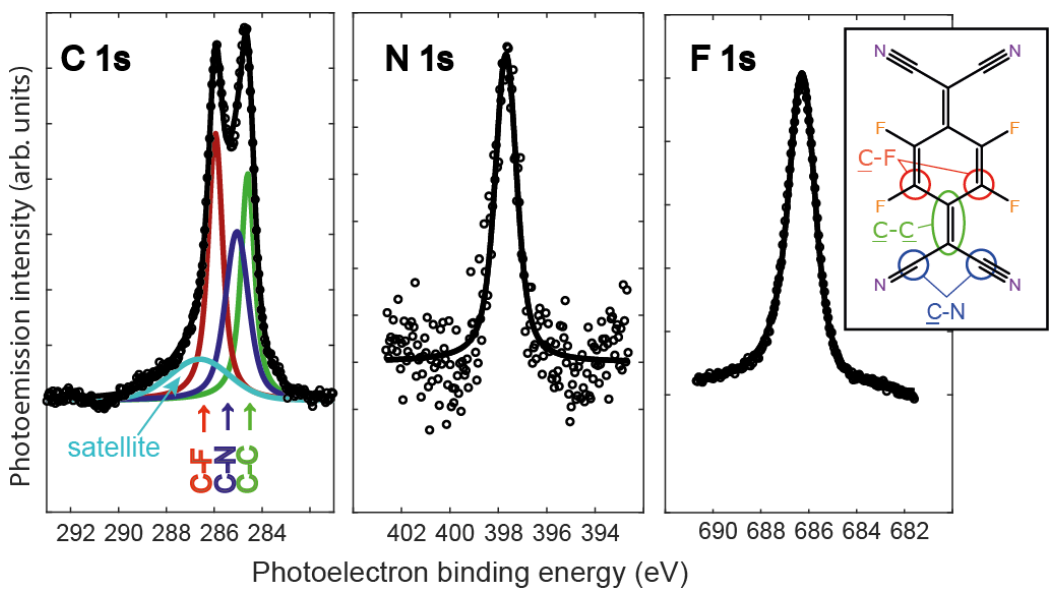

Figure 2. XPS C 1s, N 1s and F 1s spectra from the Au(111)-F$_4$TCNQ surface recorded at a photon energy of 2.2 keV (open circles), showing single components fits (continuous lines) to the N 1s and F 1s spectra and fits to the chemically shifted components of the C 1s spectrum.

NIXSW measurements employing the Au(111) reflection led to the determination of values of the two key associated structural parameters, the coherent fraction, $f$, and the coherent position, $p$. The coherent fraction is commonly regarded as an order parameter, while the coherent position is the offset of the absorber position, in units of the substrate layer spacing $d_{(111)}$, relative to the nearest extended substrate (111) plane. This can be related to the true height relative to this plane, $D = (p + n)d_{(111)}$ where $n$ is an integer chosen to ensure that interatomic distances are physically reasonable.[6] There is very rarely any ambiguity in this choice. The resulting values of $f$ and $D$ for each chemically distinct absorber atom are shown in Table 1. The experimental absorption profiles and the fits based on these values of the structural parameters are shown in Figure S1 of the Supporting Information.

**Table 1**. Summary of the coherent fractions, $f$, and coherent position values (the latter converted into a physical distance, $D$) obtained from the NIXSW measurements. Estimated precisions are shown in parentheses in units of the least significant figure. Precision estimates for $D$ in the cases of very low values of $f$ take account of the problems of achieving meaningful values of $D$ discussed elsewhere.[30]

|     | $f$       | $D$ (Å)   |
| --- | --------- | --------- |
| CF  | 0.70(10)  | 3.29(15)  |
| CC  | 0.66(10)  | 3.27(15)  |
| CN  | 0.40(10)  | 3.22(20)  |
| N   | 0.10(10)  | 2.82(40)  |
| F   | 0.37(10)  | 3.52(20)  |

As remarked above, the coherent fraction is generally regarded as an order parameter; if all absorber atoms are at the same height relative to the (111) planes, with no static or dynamic disorder, the value of $f$ would be unity. Co-occupation of two or more different heights can lead to much lower values, but even in the case of only a single height being occupied, thermal vibrations of the substrate atoms (which introduce an incoherent scattering background to the standing wave), and of the absorber atoms within the standing wave, must reduce this value by appropriate Debye-Waller factors. Careful evaluation of possible type of disorder that can occur in adsorbed molecular layers[30] leads to the conclusion that values of less than ~0.75 are likely to indicate contributions from coexisting different heights of the corresponding atomic species. The values of 0.70 for the CF atoms and 0.66 for the CC atoms fall slightly outside this limit but may be large enough to suggest that the great majority of molecules do have these C atoms at single well defined heights. By contrast, the value for the N atoms is extremely low, clearly indicating co-occupation of at least two distinctly different heights. The situation is similar, albeit less extreme, for the CN and F atoms. Precision estimates for the *D* values for these atoms include allowance for the large uncertainty arising from the low values of the coherent fractions.[30] In this context, we note that two different preparations of the adsorbate surface, both yielding the same LEED pattern, but which XPS showed to have very different average coverages of F$_4$TCNQ, did yield significantly different NIXSW coherent fractions values. Specifically, the higher coverage preparation showed low coherent fractions in the range 0.3-0.5 for all the absorber atoms; as argued elsewhere,[30] this effect can only easily be reconciled with at least partial double-layer or multilayer growth of the molecular overlayer. The lower coverage data are therefore expected to be a much better representation of the ordered single layer phase of interest, although it is possible that even in this case the slightly reduced *f* values for the CC and CF species may be due to a small fraction of the molecules occupying a second (or higher) layer.

The very much lower value of *f* for the N atoms, however, cannot be accounted for in this way; this must imply that the N atoms occupy at least two distinctly different heights above the surface, differing by up to 1 Å, in the ordered monolayer. Exactly this effect was seen in an investigation of the commensurate ordered phase of TCNQ on Ag(111),[7] and was shown to be reconcilable with a twisted molecular conformation, attributed to the presence of Ag adatoms. Specifically, the four N atoms of each molecule adsorb at two different heights, leading to the twisted molecular conformation, the upper N atoms being bonded to the adatoms while the lower N atoms are bonded to the undisturbed underlying Ag surface atoms. It therefore seems likely that a similar geometry occurs in the Au(111)-F$_4$TCNQ system if Au adatoms are involved. Notice that if N atoms occupy two or more different heights, this is also likely to occur for the CN atoms bonded to the N atoms, leading to a significant reduction in *f* for these absorbers.

The low coherent fraction of the F atoms also indicates the co-occupation of multiple heights. Higher resolution XP F 1s spectra recorded from F$_4$TCNQ adsorbed on Ag(100) [31], show the presence of a second component, weakly resolved at the higher energies of the NIXSW measurements, that appears to be related to radiation damage. This suggests that on both of these surfaces the F NIXSW results may be influenced by the presence of a coadsorbed atomic F species. In this context we note that F is known to have a particularly high cross-section for electron- and photon-stimulated desorption (*e.g.* [32]).

**Density Functional Theory Calculations**

As remarked in the introduction, the earlier STM-based investigation of the Au(111)-$F_4$-TCNQ system[9] did report results of DFT calculations for an adsorption-induced adatom structure, but the focus of these calculations was primarily on simulating STM images. No quantitative structural parameters were reported. However, a schematic side view of the adsorption structure included in this paper appears to show the molecule in the inverted bowl conformation commonly reported in DFT calculations of TCNQ adsorption on metal surfaces without adatoms being present. For the present quantitative structure determination, comparison of the calculated structural parameters with the results of the experimental NIXSW results is essential, and for this purpose the inclusion of dispersion forces is now widely recognised as being very important. We also wish, within this work, to establish from DFT calculations the energetic advantage of Au adatom incorporation into the $F_4$TCNQ overlayer, a quantity not explicitly reported in earlier studies of this system.

Our DFT calculations were performed for two alternative structural models of the $\begin{pmatrix} 5 & 2 \\ 1 & 3 \end{pmatrix}$ ordered phase, one in which there is only a single $F_4$TCNQ molecule in each unit mesh, the other in which each unit mesh contains one $F_4$TCNQ molecule and one Au adatom. The minimum energy version of each structure was then analysed to extract values for the NIXSW structural parameters that would be expected from these structures. Notice that the DFT values of the coherent fractions only take account of the reduction from the ideal value of unity due to height variations of chemically-equivalent atoms in the overlayer. No account is taken of static or dynamic disorder, so the theoretical values of the coherent fractions may be expected to be up to 30% too large for realistic comparison with experimental values[30]. Notice, too, that if an atomic species occupies two different heights that differ by one half of the bulk interlayer spacing the coherent fraction falls to zero, so the value of the coherent faction is extremely sensitive to the height difference as it approaches his value.[30]

Table 2 shows this comparison of experimental and theoretical NIXSW parameter values. For the no-adatom model the predicted heights of the C atoms in and close to the central quinoid ring are in good agreement with the experimental values, although the agreement with experiment for the heights of the other atoms is relatively poor. Most significantly, however, the no-adatom model predicts all the chemically-equivalent atoms to be at closely similar heights above the surface, resulting in high predicted values for their coherent fractions. In particular, the model completely fails to account for the exceptionally low value of the coherent fraction for the N atoms. As shown in Figure 3, in the no-adatom model the molecule adopts an inverted bowl configuration with all N atoms, at clearly similar heights, approximately 0.5 Å lower on the surface that the central quinoid ring. This molecular conformation was also found in previous DFT calculations of $F_4TCNQ$[10, 11] (and also $TCNQ$[16]) adsorbed on an unreconstructed Au(111) surface at low coverage (without the constraint of the dense packing of the low-symmetry $\begin{pmatrix} 5 & 2 \\ 1 & 3 \end{pmatrix}$ unit ,mesh).

**Table 2**. Comparison of the experimental NIXSW parameter values of Table 1 with predicted values for the optimised DFT structures of two alternative models, with and without Au adatoms. DFT atomic heights are relative to the average outermost Au layer.

|    | Expt. | DFT with adatom | DFT no adatom | Expt. | DFT with adatom | DFT no adatom |
|----|-------|-----------------|---------------|-------|-----------------|---------------|
|    | $f$   | $f$             | $f$           | $D$ (Å) | $D$ (Å)       | $D$ (Å)       |
| <u>CF</u> | 0.70(10) | 1.00 | 1.00 | 3.29(15) | 3.25 | 3.28 |
| <u>CC</u> | 0.66(10) | 0.98 | 0.98 | 3.27(15) | 3.23 | 3.15 |
| <u>CN</u> | 0.40(10) | 0.70 | 0.95 | 3.22(20) | 3.03 | 2.86 |
| N  | 0.10(10) | 0.37 | 0.82 | 2.82(40) | 2.83 | 2.62 |
| F  | 0.37(10) | 0.98 | 1.00 | 3.52(20) | 3.23 | 3.31 |

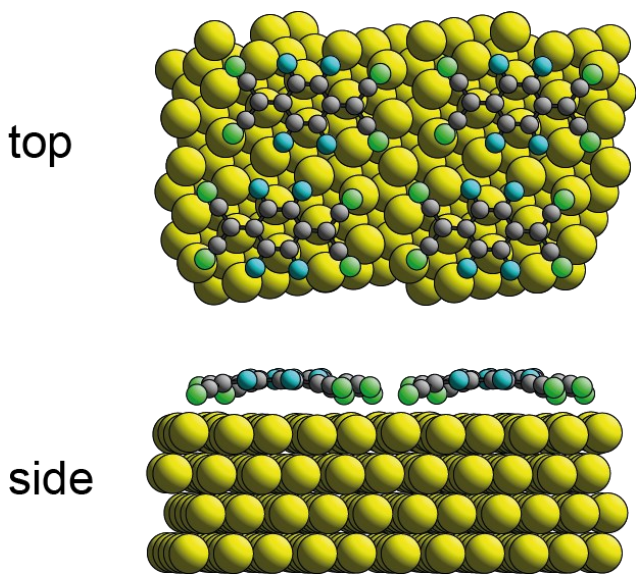

Figure 3. Top and side views of the minimum-energy DFT structure of the model based on F$_4$TCNQ adsorbed on Au(111) in a $\begin{pmatrix} 5 & 2 \\ 1 & 3 \end{pmatrix}$ unit mesh without Au adatoms. Au atoms are shown coloured yellow, C black, F blue and N green.

By contrast, the structural model including one Au adatom per surface unit mesh predicts a very low experimental coherent fraction for the N atoms. This arises because in the adatom structural model (Figures 4(a) and 4(b)) the F$_4$TCNQ molecule has twisted ends, leading to two very different heights ( by ~0.8 Å) of the N atoms (and two slightly less different heights of the C-N atoms), causing a significant reduction of the associated coherent fractions, consistent with the experimental data. The exceptionally low experimental coherent fraction for the N atoms, in particular, leads to very poor precision in the coherent position, expressed as a value of $D$. With this caveat the agreement between the experimental $D$ values for all chemically-distinct atoms and those predicted for the adatom models is good.

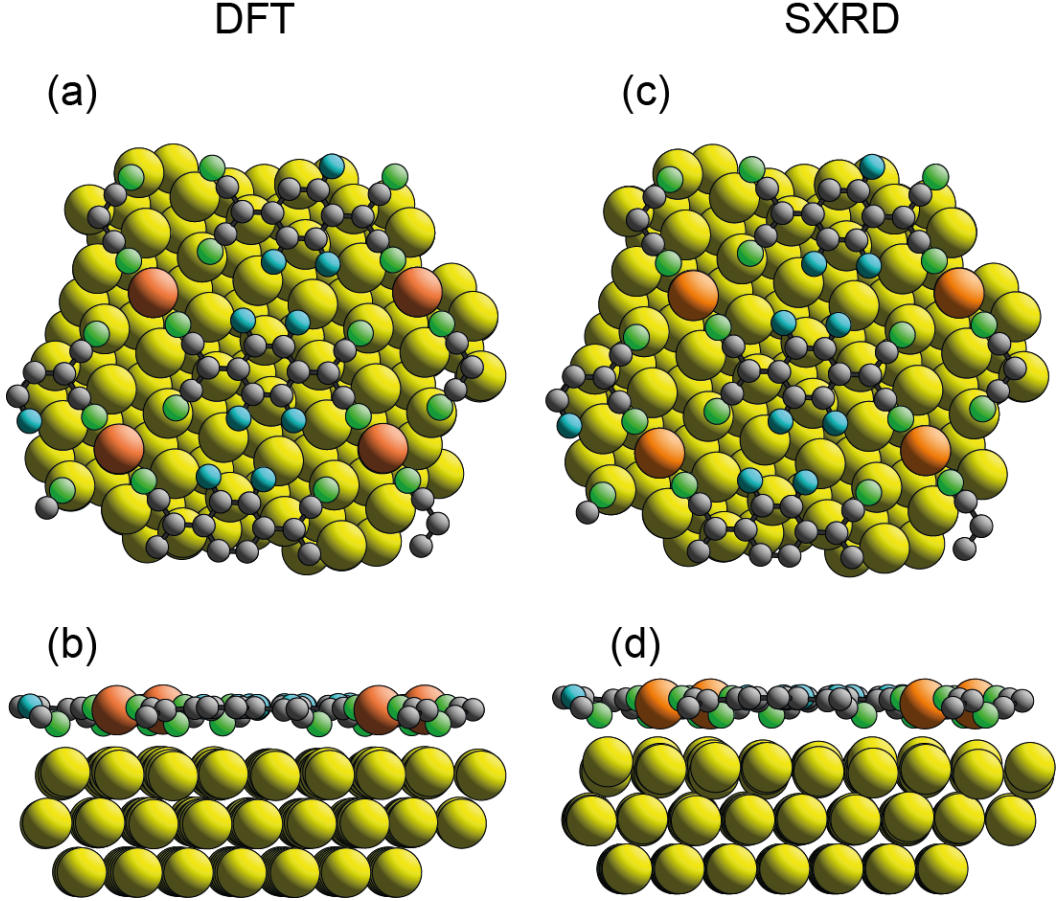

**Figure 4**. Top (a) and side (b) views of the minimum energy structure (including Au adatoms) of the $\begin{pmatrix} 5 & 2 \\ 1 & 3 \end{pmatrix}$ Au(111)-F$_4$TCNQ phase found in the DFT calculations with a superimposed unit mesh. The Au adatoms are shaded orange to distinguish them from those of the unreconstructed substrate: C atoms are shown grey, N green, and fluorine blue. (c) and (d) show the structure found in the SXRD study.

Based on the combination of NIXSW and DFT structural data there is therefore clear indirect evidence of the presence of adatoms in the Au(111)-F$_4$TCNQ $\begin{pmatrix} 5 & 2 \\ 1 & 3 \end{pmatrix}$ structure. A theoretical analysis of the energetics of the two models reinforces this conclusion. Adsorption energies ($E_{ads}$) per surface unit mesh (and thus per F$_4$TCNQ molecule) were calculated for the no adatom model (see Figure 3) as:

$$E_{ads} = E_{opt} - (E_{Au(111)} + E_{F_4TCNQ}) \qquad (1)$$

where $E_{opt}$, $E_{Au(111)}$ and $E_{F_4TCNQ}$ are the total energy of the optimised structure, of the clean Au(111) (unreconstructed) surface, and of the free molecule, respectively. For the adatom model (Figure 4(a) and (b)), equation (1) is modified to:

$$E_{ads} = E_{opt} - (E_{Au(111)} + E_{F_4TCNQ} + E_{Au} + E_{coh}) \qquad (2)$$

where $E_{Au}$ is the energy of a free Au atom (included to account for the additional Au atom per unit mesh in the adatom structure) and $E_{coh}$ is the cohesive energy of bulk Au (included to account for the energy cost of extracting the adatom). In these formulations all energies are taken to be positive. The results reveal a strong energetic advantage for the adatom model with an adsorption energy per unit surface area of 3.35 eV/nm² relative to a value of 2.57 eV/nm² in the absence of the adatom. This situation is similar to that of F4TCNQ on Ag(100)[31] for which we showed that the strong intralayer bonding in the 2D-MOF that is formed is the origin of the preference for the adatom model. The results show the optimum lateral registry of the overlayer to the underlying Au(111) surface to correspond to the Au adatoms occupying local atop sites; shifting this registry to a hollow site reduced the adsorption energy by 0.43 eV/nm², while starting in a bridge site the calculation converged on the hollow site geometry. Figures 4(a) and (b) show the minimum energy structure (which includes Au adatoms) in both plan and side views. Comparison with the results of the earlier DFT study of the adatom structure is not really possible, because the equivalent diagram shown in Figure 1 of this earlier paper[9] can only be regarded as highly schematic, showing a model with a completely different $\begin{pmatrix} 5 & 1 \\ 1 & 8 \end{pmatrix}$ unit mesh having an area 3 times larger than the true $\begin{pmatrix} 5 & 2 \\ 1 & 3 \end{pmatrix}$ unit mesh, the superimposed F4TCNQ molecule apparently being enlarged to fit this large mesh.

Comparison of the NIXSW and DFT structural results, and the DFT energetics, clearly favour the adatom model. However, these techniques do not provide any direct experimental evidence of the presence and location of Au adatoms. This information is provided by the results of the SXRD experiment described below.

**SXRD**

SXRD measurements focused predominantly on the intensities of the fractional order diffracted beams arising from the $\begin{pmatrix} 5 & 2 \\ 1 & 3 \end{pmatrix}$ unit mesh, selecting beams corresponding to a single rotational and mirror reflection domain of the complete diffraction pattern. SXRD data collection generally involves three types of measurements, namely, (i) the intensities of fractional order diffracted beams, *hkℓ* at a low value of $\ell$, referred to as 'in plane' intensities, ($\ell$ being the component of momentum transfer perpendicular to the surface); (ii) 'rod scans' of the intensities of these fractional order beams as a function of $\ell$ (known as 'fractional order rod (FOR) scans'); (iii) Rod scans of integer order beams, known as crystal truncation rod (CTR) scans. The data in (i) and (ii) are influenced only by the structure of that part of the surface that shows the periodicity of the surface phase, providing no information on the location of these atoms relative to the unreconstructed substrate. (iii) contains information on the complete structure including that of the substrate. The dataset collected from the Au(111)-$F_4$TCNQ system, identified in Figure S2, comprised mainly in-plane intensities (i) but included a small number of FORs and CTRs, restricted in their range of $\ell$ by the low photon energy (large wavelength) chosen to avoid a strong background signal of Au fluorescent X-rays.

While a full structure determination based on these data requires computer simulations derived from alternative model structures, a Patterson function map of the projection of the structure onto the surface plane can readily be produced directly from the set of in-plane measurements of the fractional order beams (*e.g.*[33]). A Patterson function map (essentially a Fourier transform of the diffracted intensities) does not show the spatial variation of the electron density directly due to the loss of phase information in these intensities (as opposed to the amplitudes), but does show a self-convolution of this quantity, and is thus a map of interatomic vectors.[34] The Patterson map obtained directly from the experimental data is shown in Figure 5(a), with the unit mesh superimposed, together with some of the dominant interatomic vectors labelled. The map is clearly dominated by one intense peak at each corner of the unit mesh. These arise from all vectors from atoms within one unit mesh to the equivalent atom in an adjacent unit mesh (including Au-Au vectors); these correspond to the primitive translation vectors of the surface mesh, so these dominant peaks show only the periodicity of the surface. However, significant structural information is provided by the features of the Patterson map within the unit mesh. These are expected to be dominated by Au-C, Au-F and Au-N vectors; intramolecular (C-C, C-F, C-N) vectors are expected to be weaker due to the smaller scattering cross-sections for these atoms, although multiple similar interatomic vectors present within the molecule may also make some of these features visible in the map. Figure 5(b) shows the main interatomic vectors expected to dominate the Patterson map for the Au adatom structure, which clearly do correspond to the main features of the experimental map. This correspondence is reinforced by Figure 5(c), which shows the Patterson map generated from calculated fractional order beam intensities for the lowest-energy adatom structural model found in the DFT calculations (Figure 4(a)). This is clearly almost identical to the Patterson map of the experimental data in Figure 5(a). By contrast, a Patterson map generated from calculated fractional order beam intensities for the same (adatom) structural model favoured by the DFT calculations, but omitting the Au adatoms, Figure 5(e), and for the lowest-energy DFT no-adatom structure (Figure 5(d)) are very different. This provides clear evidence that the features of the experimental Patterson map within the unit mesh are consistent with the Au-C, Au-N, Au-F vectors expected for the adatom structure. Thus, these results constitute a clear demonstration of the initial objective of this investigation: to show that SXRD does provide a way to 'see' metal adatoms in an unambiguous fashion. Moreover, there is clear qualitative agreement of the structure between the results of the DFT calculations and the SXRD experiments.

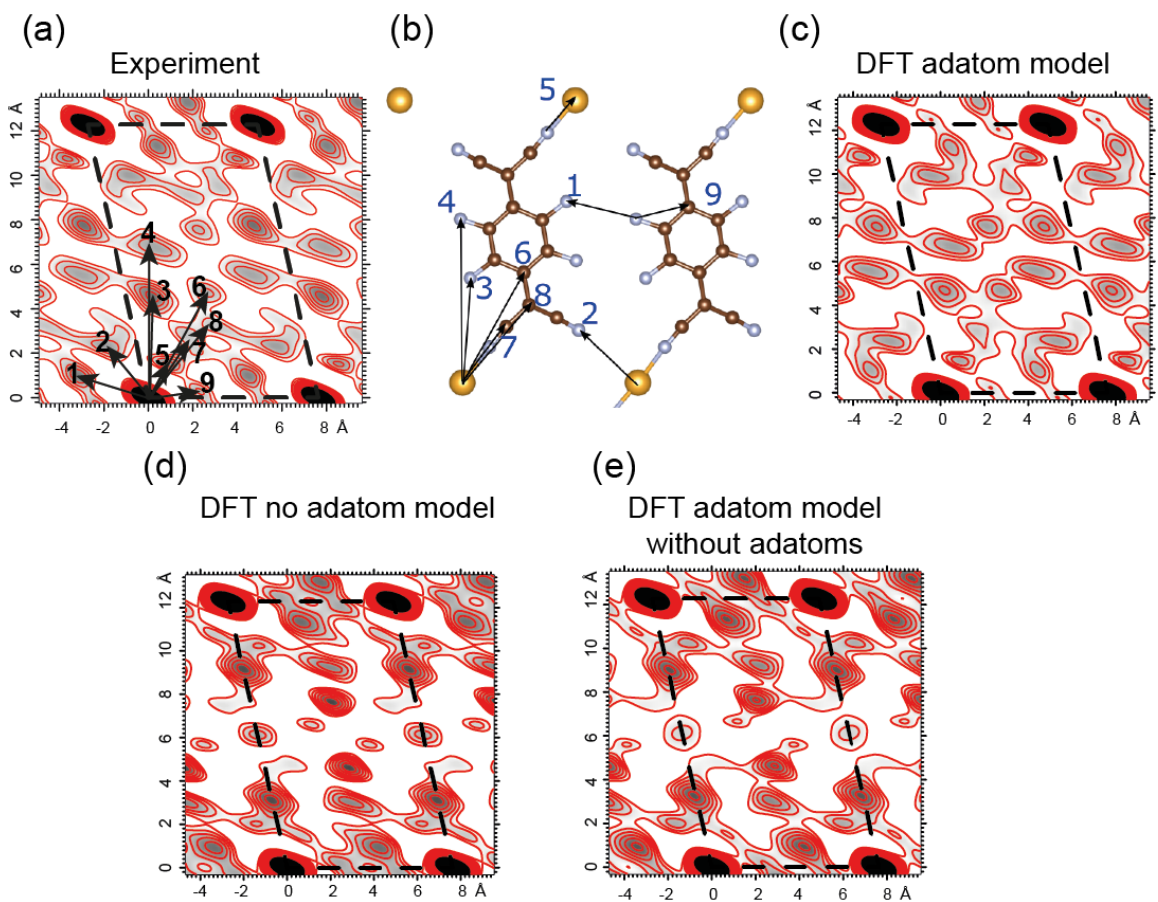

Figure 5. (a) Patterson map obtained from the experimental in-plane fractional order diffracted beam intensities with dominant numbered Au-F, Au-N and Au-C interatomic vectors superimposed. (b) shows how the interatomic vectors shown in (a) correspond to the real-space structure. Some overlapping intramolecular vectors are shown in Figure S3. (c) Patterson map produced from calculated intensities from the adatom structure found in the DFT calculations. Patterson maps based on calculated intensities from the optimal DFT structure for a no-adatom structure, and for the adatom structure with the adatoms removed, are shown in (d) and (e).

A full quantitative structure determination by SXRD is achieved by the trial-and-error approach common to almost all surface structural techniques, in which the measured quantities, in this case diffracted beam intensities, are compared with computed values to be expected for different structural models. For SXRD these computed values are provided using the ROD computer program.[35] In the present case, there is potentially a very large number of structural parameters to be determined. Specifically, these include the three Cartesian coordinates of each of the 24 constituent atoms of the $F_4TCNQ$ molecule and of the Au adatoms, but also the heights above the substrate of the 13 Au atoms per surface unit mesh in each of the outermost Au(111) layers. These layers may be rumpled as a consequence of the molecular bonding (significant rumpling in the outermost two layers is predicted by the DFT calculations). Even with our dataset comprising of 82 in-plane fractional order beam intensities, 3 FORs, and 3 CTRs, optimising all these parameter values independently is an unrealistic goal.

Unsurprisingly, exhaustive searches of models in which the coordinates of the weakly-scattering C, N and F atoms within the adsorbed molecule were varied, led to the conclusion that SXRD is too weakly dependent on these parameters to reach any conclusions about the exact molecular conformation. The small improvements in the quality of fit to the experimental data that were found often involved unphysical changes in intramolecular bond lengths and bond angles.

The details of the subsequent strategy for structural optimisation used in the trial-and-error modelling are described in the Supporting Information. This was conducted using the complete experimental dataset of in-plane and out-of-plane diffracted beam intensities identified in Figure S2. Briefly, these calculations assumed that the molecular conformation was that found in the DFT calculations. The remaining parameters to optimise were therefore the relative heights and lateral registry of the adsorbed molecule and the Au adatoms, but also the layer spacings and rumpling of the outermost Au surface layers; this rumpling proved to be important to achieve the best agreement between theory and experiment.

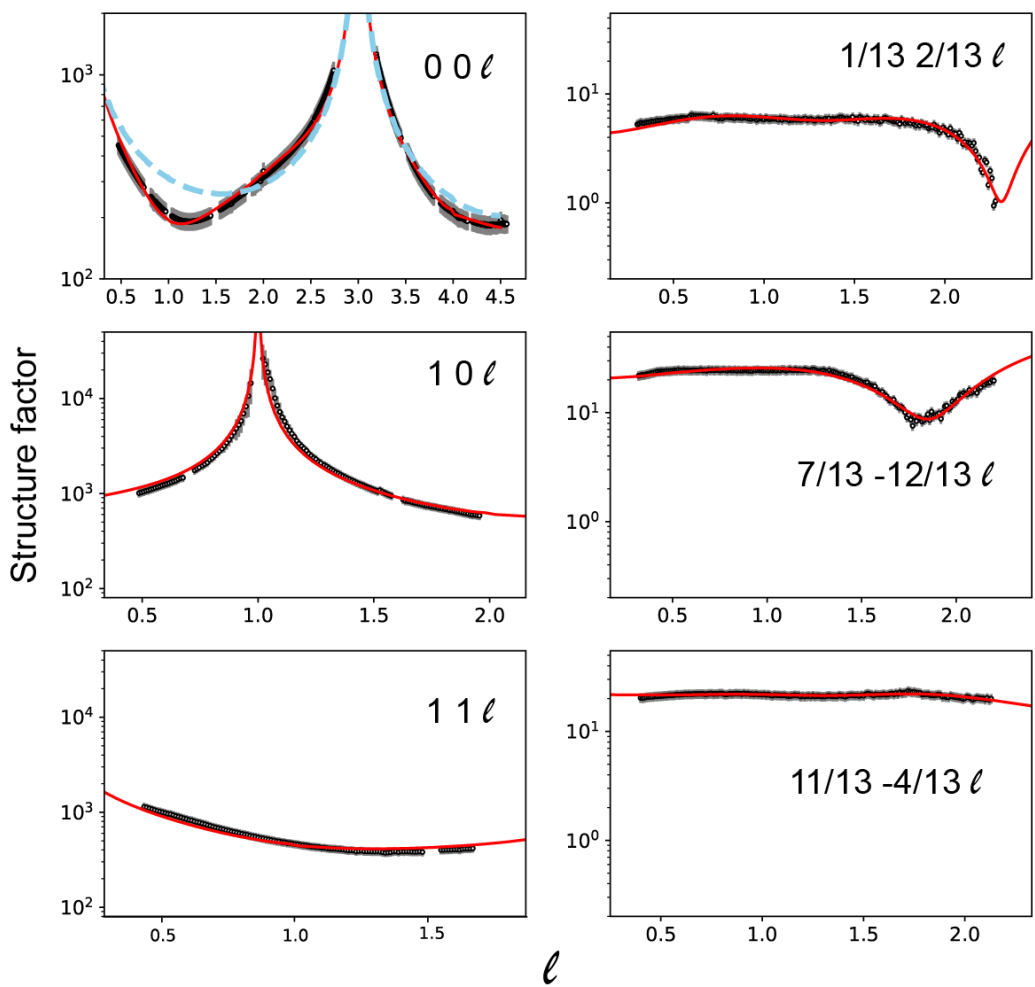

Figure 6. Experimental rod scans (individual data points shown in black with error bars) compared with the results of the ROD simulation for the optimised structural model (red continuous lines). The blue dashed curve in the 0 0 $\ell$ panel corresponds to the results of a calculation from a clean unrelaxed Au(111) surface.

Figure 6 shows a comparison the experimental rod scans with the results of calculations for the best-fit structure, while the comparison of the experimental and computed in-plane data are shown in Figure S4. The agreement is clearly good. Table 3 summarises the overlayer spacing found in the SXRD analysis, compared with the equivalent values from the DFT calculations and the NIXSW experimental results. Note that as the molecular conformation in the SXRD calculations was assumed to be that of the DFT calculations, reporting separate values for the heights of the constituent atoms in this comparison is not really meaningful, so only the heights of the central quinoid ring are compared in this table. The SXRD precision estimates for the molecule and adatom heights are determined by the range of individual parameter values that lead to a value of chi-squared within 5% of the best-fit value; there was some evidence of coupling between the values of the adatom height and the rumpling amplitudes, so the error estimates taking account of this may be increased to about ±0.10 Å. Clearly, all three methods agree within these estimated experimental precisions. A key conclusion is therefore not only that the SXRD results show that the Au adatoms do exist in the overlayer, but also that their height above the surface is determined to be consistent with the DFT model. In detail, the Au adatoms are found in the SXRD structure to be 0.27±0.06 Å below the molecular layer, to be compared with 0.14 Å in the DFT model. One further important finding, however, is that the SXRD fit shows the amplitude of the rumpling of the outermost Au layer, in particular, to be significantly larger than that indicated by the DFT calculations. Specifically, the SXRD results show the rumpling amplitudes of the outermost and second Au(111) layers to be 0.60±0.06 Å and 0.18±0.07 Å, respectively, to be compared with values of 0.20 Å and 0.09 Å, respectively, in the DFT model. The rumpling amplitude is defined here as the difference in height of the highest and the lowest atoms within the layer, but the particularly large value of this amplitude found in the SXRD analysis for the outermost layer is attributable to the two Au atoms that lie directly below the N atoms not bonded to Au adatoms, which lie 0.60 Å above the lowest top layer atoms. As these Au atoms are forming Au-N bonds to the molecule, a larger outward displacement of these atoms is qualitatively reasonable, but the large magnitude of this effect is difficult to reconcile with a shift of these atoms of only 0.1 Å in the DFT results. The rumpling of the remainder of the outermost layer atoms in the SXRD model is 0.38 Å.

**Table 3** Comparison of the height of the central quinoid ring, and of the Au adatom, above the average outermost Au layer obtained from NIXSW, SXRD and the DFT calculations for the adatom model of the $\begin{pmatrix} 5 & 2 \\ 1 & 3 \end{pmatrix}$ Au(111)-F$_4$TCNQ phase.

| atom | DFT (with adatom) $D$ (Å) | NIXSW $D$ (Å) | SXRD $D$ (Å) |
|---|---|---|---|
| C quinoid | 3.24 | 3.29±0.15 | 3.36±0.05 |
| Au adatom | 3.10 | | 3.09±0.04 |

The fact that the SXRD analysis indicates significantly enhanced surface layer rumpling, relative to that predicted by the DFT calculations, highlights another potentially important feature of the SXRD technique. Experimental data on this rumpling effect is rather scarce because most surface structural techniques used to investigate molecular adsorption structures focus on the relative location of atoms in the adsorbate and are 'blind' to displacements of the substrate atoms. The notable exception is quantitative LEED (QLEED), also known as LEED I-V (intensity-voltage) analysis, which is closely similar to SXRD but with X-rays replaced by low energy electrons. This technique[36] has identified substrate surface rumpling in atomic adsorption and adsorption of diatomic molecules such as CO, but is extremely challenging to apply to large surface mesh structures associated with adsorption of larger molecules because the importance of multiple scattering leads to much higher computational demands than SXRD. The origin of the difference in rumpling amplitude between the SXRD structure and the DFT result is unclear but may be related to the limited number of Au atomic layers in the DFT 'slab', due to the computational demands of this large surface mesh structure. However, it would be surprising if this could account for such a large difference in the rumpling amplitude of the outermost layer.

Fits to the SXRD data with the lateral registry of the Au adatoms constrained to the four alternative high-symmetry sites of the outermost Au layer clearly support the same atop site as the DFT calculations. The lowest chi-squared value, 1.236, was found for this atop site, whereas the values for the alternative registries were in the range of 1.35-1.69. Table

S1 shows the specific values and structural parameter values for each alternative registry. Notice that the Au adatom-Au surface atom nearest neighbour distance in this atop registry is 2.99 Å according to the DFT model, which is significantly larger than the Au-Au spacing in the bulk crystal (2.88 Å), suggesting that there is no Au-adatom-Au surface atom bonding. This would imply that the height of the Au adatom above the surface is determined more by the molecule-substrate and molecule-adatom bonding, than by any adatom-substrate bonding. This behaviour would be similar to that seen in the two-dimensional metal-organic framework formed by K coordinated to TCNQ on the Ag(111) surface, in which the K atoms occupy off-atop sites.[37] The value of this Au-Au adatom spacing obtained from the SXRD analysis of 3.06±0.06 Å is even larger. As a more quantitative test of the sensitivity of the SXRD data to the presence of the Au adatoms, further ROD calculations were performed starting from the DFT no-adatom model and using the same strategy as for the adatom model to search for modifications of this structure that yielded the best fit to the experimental data. The chi-squared value for the resulting model was 6.611, to be compared with the value of 1.236 for the adatom model. This model gave a particularly poor fit to the in-plane fractional order intensities as shown in Figure S6, with almost all predicted structure values very significantly too low, consistent with the absence of the strongly-scattering Au adatoms..

**CONCLUSIONS**

Despite several reports of indirect evidence of molecular adsorption on noble metal surfaces, leading to the presence of metal adatoms that are incorporated into the molecular overlayer, there has been no positive experimental identification by a quantitative structural technique. The main evidence has come from interpretation of STM images that, even aided by simulated images based on DFT calculations and the use of the Tersoff-Hamman approach, are *not* unambiguous. By contrast, X-ray diffraction is an extremely well-established technique, the theory of which is well understood, and has been used to solve crystal structures of great complexity, notably in macromolecular chemistry and the life sciences. Here we have shown that *surface* X-ray diffraction can be used to provide the necessary unambiguous identification of Au adatom creation following the adsorption of F$_4$TCNQ on Au(111). While this adsorption system was identified as a likely example of this adatom incorporation by an earlier combined STM/DFT study, our new study has clarified many important aspects of this model system, including: (i) Experimental determination of the height of the adsorbed molecule above the surface using NIXSW; (ii) Evidence from NIXSW experimental data that the adsorbed molecule is twisted and does not adopt an inverted bowl configuration as previously proposed; (iii) Application of dispersion-corrected DFT calculations to determine the preferred adsorption geometry, which is in excellent agreement with the experimental NIXSW results and the conclusions regarding the molecular conformation; (iv) Use of dispersion corrected DFT calculations to quantify the strong energetic advantage of the adatom incorporation structure over adsorption on an unreconstructed surface; (v) Demonstration that experimental surface X-ray diffraction unambiguously shows the presence of Au adatoms in the adsorption structure, with the molecule and adatom heights above the surface, and the lateral registry of the overlayer, consistent with the DFT results; (vi) The importance of rumpling in the outer Au layers.

Clearly, these results show that SXRD can be used to explore the phenomenon of adsorption-induced adatom creation in a range of other systems and has proved to be a crucial complementary experimental technique to achieve a better understanding of this effect.

**Conflicts of interest**

The authors declare no conflicts of interest.

**ASSOCIATED CONTENT**

**Supporting Information**

NIXSW results; SXRD structure analysis details

**AUTHOR INFORMATION**

**Corresponding authors**

David Phillip Woodruff − Department of Physics, University of Warwick, Coventry CV4 7AL, U.K.; orcid.org/0000-0001-8541-9721; Email: d.p.woodruff@warwick.ac.uk

**ACKNOWLEDGEMENTS**

The authors thank Diamond Light Source for allocations SI14884-1 and SI4884-2 of beam time at beamline I09 and SI19105 at I07 that contributed to the results presented here. P.J.B. and P.T.P.R. acknowledge financial support from Diamond Light Source and EPSRC. G.C. acknowledges financial support from the EU through the ERC Grant "VISUAL-MS" (Project ID: 308115). B.S. and R.J.M. acknowledge doctoral studentship funding from the EPSRC and the National Productivity Investment Fund (NPIF). R.J.M. acknowledges financial support *via* a UKRI Future Leaders Fellowship (MR/S016023/1). We acknowledge computing resources provided by the EPSRC-funded HPC Midlands+ Computing Centre (EP/P020232/1) and the EPSRC-funded Materials Chemistry Consortium (EP/R029431/1) for the ARCHER2 U.K. National Supercomputing Service (http://www.archer2.ac.uk).

TOC graphic

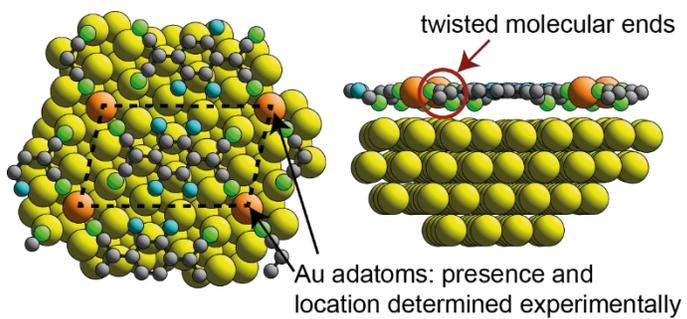

# Supporting Information
# Direct Experimental Evidence for Substrate Adatom Incorporation into a Molecular Overlayer

Philip J. Mousley[1], Luke A. Rochford[2], Paul T.P. Ryan[1,3], Philip Blowey[1,4], James Lawrence[5], David A. Duncan[1], Hadeel Hussain[1], Billal Sohail[5], Tien-Lin Lee[1], Gavin R. Bell[4], Giovanni Costantini[5], Reinhard J. Maurer[5], Christopher Nicklin[1], D. Phil Woodruff[4]*

[1]*Diamond Light Source, Harwell Science and Innovation Campus, Didcot, OX11 0DE,*
[2]*Chemistry Department, University of Birmingham, University Road, Birmingham B15 2TT, UK*
[3]*Department of Materials, Imperial College, London SW7 2AZ, UK*
[4]*Department of Physics, University of Warwick, Coventry CV4 7AL, UK*
[5]*Department of Chemistry, University of Warwick, Coventry CV4 7AL, UK*

1. **NIXSW results**

2. **SXRD structure analysis details**

1. **NIXSW results**

Figure S1 shows a comparison of the raw NIXSW photoemission intensity scans with the best-fit theoretical curves, the two fitting parameters (coherent fraction and coherent position) being reported in Table 1 of the main paper. Values of the backward-forward asymmetry factor, $Q$, were 0.11 for the C and N emitters and 0.10 for the F emitter.

---

* Email: d.p.woodruff@warwick.ac.uk

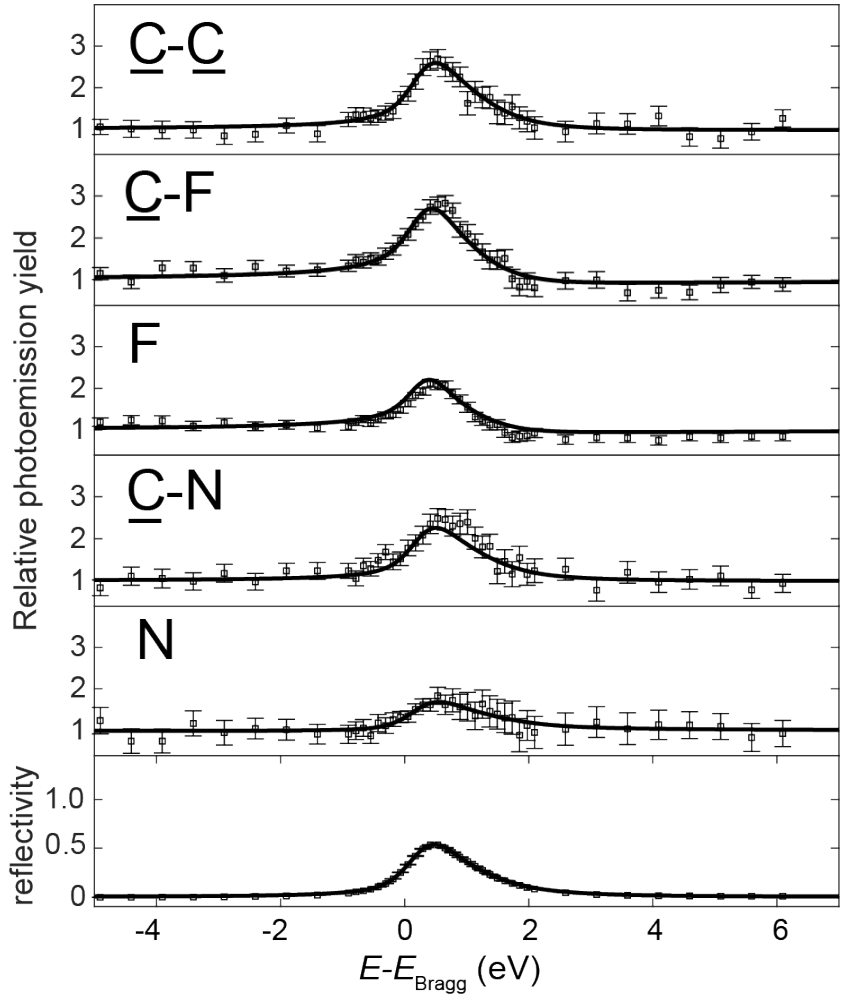

**Figure S1** Individual experimental chemical-state-specific NIXSW scans (data points) with the superimposed fits corresponding to the values of the coherent fractions and position reported in Table 1 of the main paper.

## 2. SXRD structure analysis details

The complete structural data set measured from a single rotational/mirror domain of the Au(111)-F$_4$TCNQ adsorption phase comprised 82 fractional-order beam 'in plane' intensities, three CTR scans ( (00$\ell$), (10 $\ell$) and (11$\ell$)) and three FOR scans ((1/13, 2/13 $\ell$), (7/13 -12/13 $\ell$) and (11/13 -4/13 $\ell$)), this labelling being based on a $\begin{pmatrix} 2 & -1 \\ 3 & 5 \end{pmatrix}$ matrix; Figure S2 specifies these.

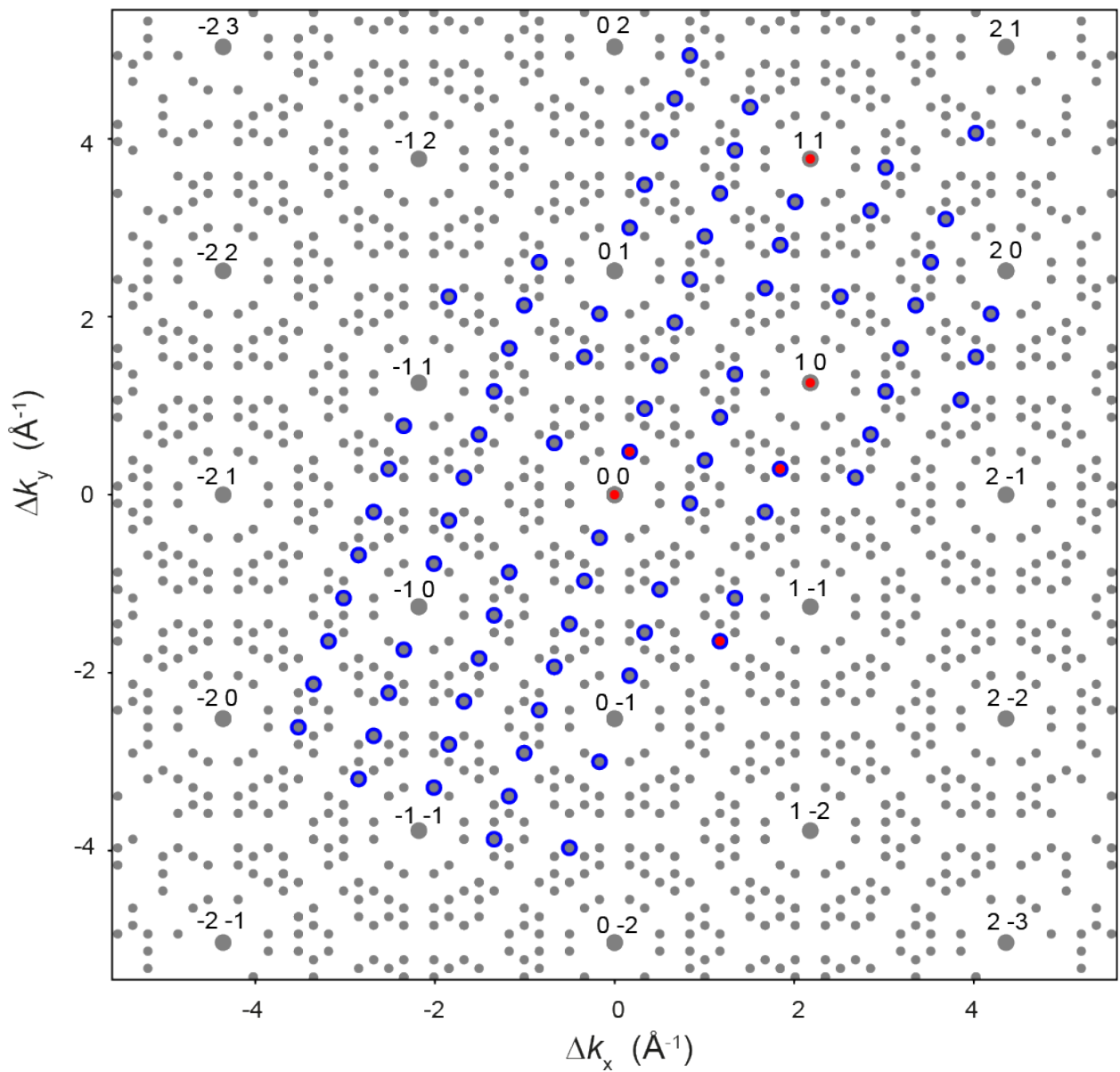

**Figure S2** Representation of the diffraction pattern of the Au(111)-F$_4$TCNQ adsorption phase from all coexistent rotationally and mirror reflected domains, identifying the SXRD dataset measured from a single such domain of the complete pattern. In-plane intensities were measured from the beams circled in blue, while rod scans were obtained from those marked with red circles. Integral order beams are labelled.

In principle, 'in plane' intensities correspond to ($hk$0) beams – i.e. to measurements with the perpendicular momentum transfer $\ell$ equal to zero. In reality, measurements are not possible at $\ell$ =0, but can be obtained at small values of $\ell$. In the present case these 'in plane' intensities were measured at $\ell = 0.2$. Friedel's rule tells us that the intensity of the ($-h\ -k\ell$) beam is the same as an ($hk\ -\ell$) beam, so an average of the intensities of an ($-h\ -k\ell$) beam and the corresponding ($hk\ell$) beam at low values of $\ell$ can provide a good interpolated estimate of the true in-plane ($hk0$) intensity. However, for the dataset collected, calculations of the $\ell$ dependence of the fractional order beam intensities based on the DFT structrural model indicated that this linear interpolation approach could lead to intensity errors of ~10% or more in some cases, so all theory-experiment calculations were based on the true measured values of $\ell$, based on a symmetry of P1; thus, no symmetry averaging was applied.

The Patterson map based on these in-plane measurements is shown in Figure 5 of the main paper; included in this figure as part 5(b) is a real-space representation of the structure showing some of the dominant interactomic vectors, mostly involving Au adatoms. Figure S3 shows an extended set of interatomic vectors superimposed on a model of the real space structure covering 4 unit meshes, also including some multiply-occuring inter- and intra-molecular vectors.

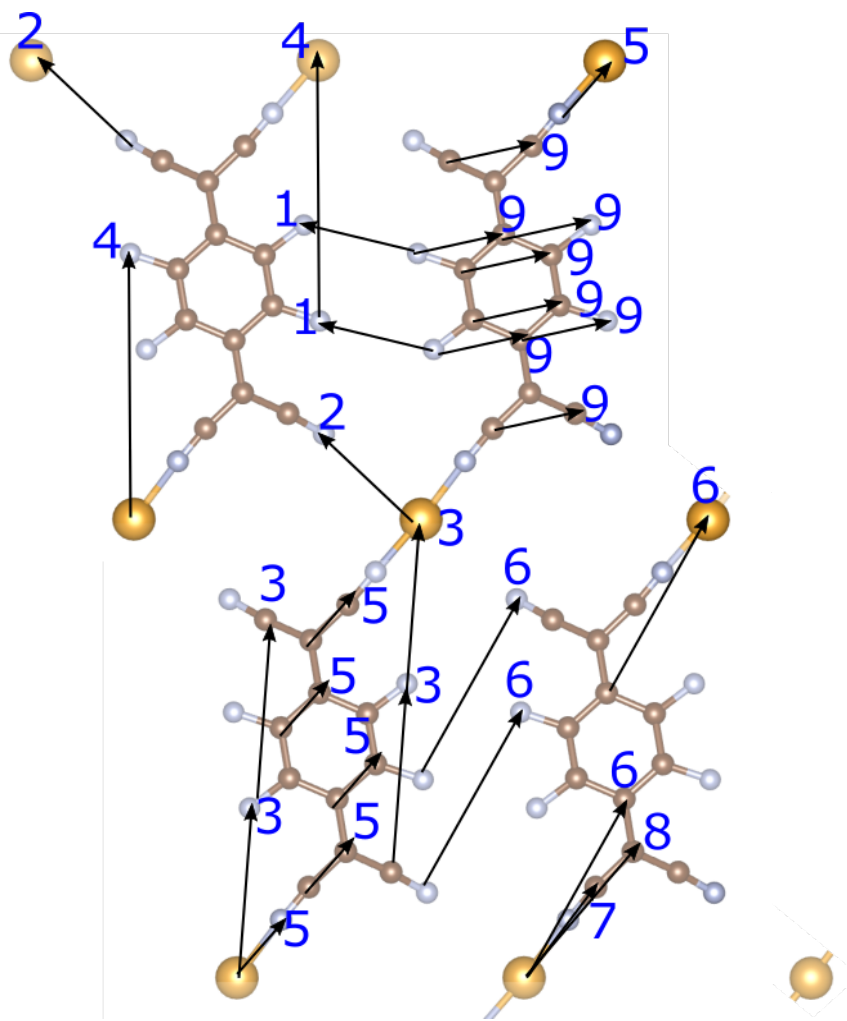

**Figure S3** Diagram of the surface structure over four unit meshes showing the repeat occurrences of in-plane vectors labelled 1-9, identified in the Patterson map shown in Figure 5(a) of the main paper. The number at the head of each arrow corresponds to its vector label.

Full quantitative structure determination using SXRD relies on the trial-and-error fitting process common to almost all surface structural technques, the experimental diffracted beam intensities being compared with those predicted in simulations based on a series of alternative structural models. For this purpose, calculations used the ROD computer program.[1] Initially the parameter search was performed on models based on a planar $F_4TCNQ$ molecule together with an Au adatom placed at different heights and lateral positions on a bulk-terminated substrate, adjusting a range of parameters defining distortions of the molecular conformation including in-plane and out-of-plane displacements of individual atoms. The experimental dataset used for this initial theory/experiment comparisons was the complete set of 'in plane' fractional-order beam intensities and the rod scan of just one fractional order beam, namely (1/13 2/13 $\ell$). The main conclusion of exhaustive structural searches based on this approach was that the experiental data were too insensitive to detailed parameters defining the molecular conformation to reach any conclusions on this aspect of the structure. This conclusion is, of course, consistent with the known weak scattering cross-sections of the atomic constituents of the molecule. Further structural searches therefore assumed the molecular conformation was as given by the DFT calculations; this choice was reinforced by the fact that the relative heights of the constituents atoms in this model are consistent with the NIXSW experimental results.

A second stage of the structure determination focussed on the (00) CTR with the objective of identifying the various layer spacings (of the molecule and adatom, but also of any relaxation of the outermost substrate layer spacings), an approach based on the recognition that this specular reflectivity scan is sensitive *only* to these layer spacings. These calculations showed that the fit to the (00) rod is extremely sensitive to the spacings of the outermost substrate layers (with changes of only 0.01 Å leading to detectable changes), but far less sensitive to the heights of the molecule and Au adatom, changes of up to ~0 2 Å in these parameters leading to changes within the precision limits of the expetimental measurements. This difference in sensitivity reflects the difference in scattering cross-sections of complete Au(111) layers (each containing 13 Au atoms per surface unit mesh) and layers containing only one Au adatom or one complete molecule (comprising low atomic number atoms). Despite the relatively weak sensitivity to the exact location of these layers, a good fit to the data could only be achieved with both the molecule and the adatom being present in the structural model. The (00) CTR also proved to be sensitive to changes in the rumpling amplitude of the outermost most layers.

In the third stage of the structural optimisation, the dataset used for comparison was enlarged to include the full set of three fractional order rod scans and three integral order CTRs (see Figure S3), as well as the complete in-plane dataset. This revealed a strong sensitivity, particularly for the fractional order rods, to the surface layer rumpling amplitudes. Modifying the specifics of the DFT rumpling (i.e. which surface atoms within the surface unit mesh were displaced by how much), or refining the details of rumpling added to a model with initially planar surface layers, did not significantly alter the quality of the experiment-theory fits, but variations in the amplitude of the rumpling did prove to be important. Initially, the effect of modifying the details of the rumpling of the DFT model was explored. the final oprimsed structure was obtained by starting with planar outer layers and allowing the full range of displacements perpendicular to the surface of all the independent Au atoms in the two outermost layers.

The optimised surface structure obtained using this approach, which yielded a chi-squared value of 1.236, is illustrated in Figures 4(c) and (d) of the main paper, with the fits to the rod scans shown in Figure 6, while the fit to the in-plane data for this model is shown in Figure S4. The adatom and molecule layer spacings for this structure are listed in Table 3 of the main paper.

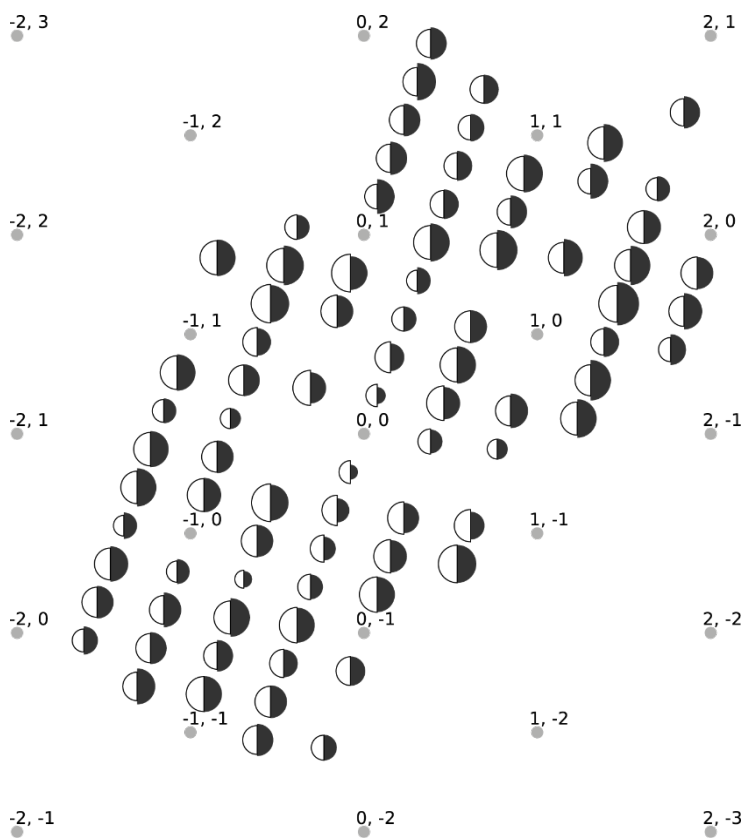

**Figure S4** Comparison of the experimental (black semicircles) and predicted (open semicircles) in-plane structure factors of the full set of fractional order beams for the optimised SXRD structure shown in Figures 3(c) and (d) of the main paper. The areas of the semicircles are proportional to the measured and calculated structure factors

As described briefly in the main text, the influence of alternative lateral registries of the overlayer, relative to the outermost Au(111) layer, was explored, re-optimising the structural parameters for the best fit to the SXRD data for the Au adatom in atop, fcc hollow (atop third layer Au atoms), bridge and hcp hollow (atop second layer Au atoms). The chi squared values and the associated layer spacings for each registry are shown in Table S1, showing a clear preference for the Au adatom to occupy a local atop site relative to the outermost Au(111) layer.

Figure S6 shows a similar comparison of the measured in-plane structure factors with the results of ROD calculations for the best-fit no-adatom model, starting from the DFT no-adatom model but refining the structure in the same way as for the adatom model. The agreement is clearly poor, with predicted structure factors almost all too small. As reported in the main paper the chi-squared value for this model (for the complete experimental dataset of in-plane and rod scan measurements) was 6.611.

**Table S1** Comparison of the chi-squared values and structural parameter values for fits to the SXRD data based on different lateral registry sites of the Au adatoms.

| Adatom registry | Chi squared | Molecule height (Å) | Adatom height (Å) | Top layer rumpling amplitude (Å) | Second layer rumpling amplitude (Å) |
|---|---|---|---|---|---|
| atop | 1.236 | 3.36 | 3.09 | 0.60 | 0.18 |
| fcc hollow | 1.350 | 3.12 | 3.02 | 0.58 | 0.21 |
| bridge | 1.691 | 3.45 | 1.76 | 0.55 | 0.22 |
| hcp hollow | 1.597 | 3.10 | 3.11 | 0.64 | 0.26 |

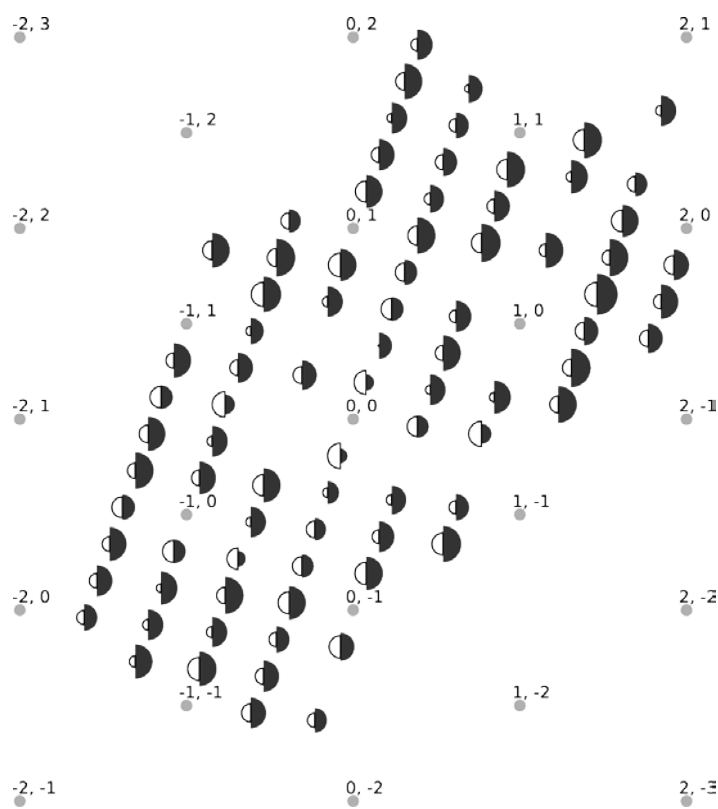

**Figure S6** Comparison of the experimental (black semicircles) and predicted (open semicircles) in-plane structure factors of the full set of fractional order beams for the optimised no-adatom SXRD structure. The areas of the semicircles are proportional to the measured and calculated structure factors.

---

[1] Vlieg, E. ROD: a Program for Surface X-Ray Crystallography, *J. Appl. Cryst.* **2000**, *33*, 401-405.